\documentclass[runningheads]{llncs}
\usepackage[utf8]{inputenc} \usepackage[T1]{fontenc}
\usepackage{booktabs}
\usepackage{array}
\usepackage{multirow}
\usepackage{graphicx}
\usepackage{float}
\usepackage{xcolor}
\providecommand\IfDocumentMetadataT[1]{}
\PassOptionsToPackage{hyphens}{url}
\usepackage{hyperref}
\usepackage{amsmath,amssymb}
\usepackage{url}
\usepackage{algorithm}
\usepackage{algorithmic}
\usepackage{listings}
\usepackage{subcaption}
\usepackage{tabularx}
\let\origunderscore\_

\usepackage{tikz}
\usetikzlibrary{positioning,arrows.meta,fit,shapes.geometric,calc}
\usepackage{pgfplots}
\pgfplotsset{compat=1.16}
\definecolor{phishred}{RGB}{190,30,45}
\definecolor{legitgreen}{RGB}{38,139,78}
\definecolor{accentblue}{RGB}{38,80,160}
\lstdefinestyle{emailstyle}{
basicstyle=\ttfamily\footnotesize, frame=single, columns=fullflexible, keepspaces=true, breaklines=true, showstringspaces=false }
\hypersetup{colorlinks=true,linkcolor=accentblue,citecolor=accentblue,urlcolor=accentblue}
\newcommand{\dsTotalRecords}{29{,}142}
\newcommand{\dsPhishCount}{14{,}571}
\newcommand{\dsLegitCount}{14{,}571}

\newcommand{\numSeeds}{5}
\newcommand{\totalRuns}{25}
\newcommand{\entityCoverageAll}{100.0\%}
\newcommand{\headerCoverageFrom}{100.00\%}
\newcommand{\headerCoverageSubject}{100.00\%}
\newcommand{\headerCoverageMessageId}{100.00\%}
\newcommand{\avgEntityTypes}{14.9}
\newcommand{\entityLabels}{40}

\newcommand{\numRelations}{5}
\newcommand{\featureDim}{1{,}600}
\newcommand{\psoParticles}{30}
\newcommand{\psoIters}{50}
\newcommand{\psoSparsityLambda}{0.015}
\newcommand{\engineThroughput}{60}
\newcommand{\totalTimeMin}{148.4}
\newcommand{\numTacticPatterns}{41}
\newcommand{\ruleIocFOne}{0.2031}
\newcommand{\ruleIocPrec}{0.6830}
\newcommand{\ruleIocRec}{0.1193}
\newcommand{\spamBayesFOne}{0.9583}
\newcommand{\spamBayesPrec}{0.9967}
\newcommand{\spamBayesRec}{0.9227}
\newcommand{\ppFOne}{0.9675}
\newcommand{\ppPrec}{0.9736}
\newcommand{\ppRec}{0.9614}
\newcommand{\psoJaccard}{0.960}
\newcommand{\numUniqueTactics}{41}
\newcommand{\numNovelTactics}{32}
\newcommand{\pctNovelTactics}{78.0}
\newcommand{\numMitreCovered}{11}

\newcommand{\ppFOneAdvHigh}{0.9579}
\newcommand{\spamBayesAdvHigh}{0.0243}
\newcommand{\distilBertAdvHigh}{0.7284}
\newcommand{\maxTrainPSO}{500}

\title{PhishSigma++: Malicious Email Detection with Typed Entity Relations}
\titlerunning{PhishSigma++: Malicious Email Detection}
\author{
Shang Shang\inst{1,2}\thanks{This work was supported by the National Key R\&D Program of China (No. 2024YFB3109004), the National Natural Science Foundation of China (No. 62202466), the Youth Innovation Promotion Association CAS (No. 2022159), the Key Laboratory of Network Assessment Technology, Chinese Academy of Sciences, and the Beijing Key Laboratory of Network Security and Protection Technology.}
\and Ruiqi Wang\inst{1}
\and Ruijie Qi\inst{1,2}
\and Hao Li\inst{1,2}
\and Yingxiao Xiang\inst{1}
\and Yepeng Yao\inst{1,2}
\and Zhengwei Jiang\inst{1,2}}
\authorrunning{S. Shang et al.}
\institute{Institute of Information Engineering, Chinese Academy of Sciences, Beijing, China\\
Corresponding author: Ruiqi Wang.\ \
\email{\{wangruiqi, qiruijie, lihao, xiangyingxiao, yaoyepeng, jiangzhengwei\}@iie.ac.cn}
\and
School of Cyber Security, University of Chinese Academy of Sciences, Beijing, China\\
\email{shangshang20@mails.ucas.ac.cn}}
\date{}
\begin{document}
\maketitle
\begin{abstract}
With the rise of AI-generated content (AIGC), phishing actors now possess richer linguistic capabilities and a wider range of evasion techniques. Our analysis shows that most existing detectors over-rely on mutable textual features. As a result, they achieve high accuracy on clean datasets yet exhibit severe degradation under text-focused adversarial manipulation. This mirrors the performance gap widely observed between laboratory benchmarks and real-world deployments. To address this discrepancy, we investigate invariant signals in phishing emails and observe that, even when attackers freely modify surface text, the functional intent of the attack constrains relations among certain typed entities. Although threat-actor tradecraft is often described through high-level tactics, techniques, and procedures (TTPs), rule-based systems such as Sigma express these invariants only through manually curated, field-specific literal patterns. This limits flexibility and coverage.
We introduce \emph{PhishSigma++}, an entity–relation–based malicious email detector for RFC822 messages that generalizes the design philosophy of Sigma rules. The system extracts \entityLabels{} typed entity classes, computes \numRelations{} cross-type relations to construct a typed email graph, and employs particle swarm optimization (PSO) to select a sparse discriminative mask. This mask supports both classification and a type-level evidence summary, enabling auditable reasoning over cross-field invariants. On a corpus of \dsTotalRecords{} messages, PhishSigma++ achieves \ppFOne{} F$_1$ on clean data and substantially outperforms text-centric baselines under non-adaptive Good Word padding at $\rho = 0.8$. It maintains \ppFOneAdvHigh{} F$_1$, while a token-based Bayesian filter collapses to \spamBayesAdvHigh{} and a released DistilBERT phishing-email checkpoint falls to \distilBertAdvHigh{}. Compared with traditional Sigma rules, PhishSigma++ provides higher detection performance, broader coverage of relational invariants, and data-driven feature selection. We further show that thresholded typed relation scores induced by our relation functions encode a useful fragment of Sigma-style field conditions, placing hand-crafted rule logic and learned relation masks in one single-email representational framework.
\keywords{Malicious email detection \and Typed entity relations \and Adversarial robustness \and Sigma rules \and Phishing}
\end{abstract}
\section{Introduction}\label{sec:intro}
Recent advances in generative models have significantly expanded the attacker's toolbox in phishing and other forms of malicious email campaigns. With large language models able to generate fluent and highly customized text at scale, attackers can now diversify email wording at negligible cost and conceal malicious intent through sophisticated paraphrasing, narrative restructuring, or content padding. These capabilities make contemporary phishing messages increasingly resistant to detectors that rely heavily on textual surface features.
Our analysis shows that many existing machine learning based email defenses exhibit an implicit dependence on mutable body text. This reliance results in strong performance on clean benchmark datasets but substantial degradation when confronted with adversarial manipulations that specifically target textual features. The gap between clean and adversarial performance helps explain why classifiers that appear highly accurate in controlled evaluations often underperform in real deployment scenarios~\cite{lowd2005good}.
To address this problem, we observe that although an attacker can freely manipulate linguistic content, a phishing email must still preserve certain structural and semantic relationships among key entities in order to achieve its objective. These relationships include stable interactions among sender identity, displayed and actual link targets, brand references, role semantics, and other elements that together define the social engineering tactic. In other words, many tactics, techniques, and procedures constitute higher‑level invariants that cannot be simultaneously altered without undermining the attack's purpose.
Sigma rules embody a similar philosophy by encoding such invariants in manually curated field conditions. However, Sigma rules require human engineering effort, lack flexibility, and do not generalize well beyond their handwritten patterns~\cite{sigma2023spec,gao2021enabling}. Motivated by these limitations, we introduce \emph{PhishSigma++}, an entity–relation based malicious email detector for RFC822 messages. PhishSigma++ automatically extracts typed entities, computes pairwise relations between them, and identifies a sparse subset of discriminative relations through an optimization‑guided mask. This formulation extends the core idea behind Sigma‑style field consistency rules into a more expressive and data‑driven framework that remains auditable at the type level.
Our main contributions are as follows:
\begin{itemize}
  \setlength{\itemsep}{0.2em}
  \setlength{\topsep}{0.2em}
  \setlength{\parsep}{0pt}
    \item We propose a learned relation-masking mechanism that automatically identifies a sparse set of discriminative entity‑relation patterns. Compared with manually curated Sigma rules, the data‑driven mask yields consistent detection improvements in our experiments without relying on brittle textual surface features.
    \item We formalize how thresholded typed relation scores induced by the collapsed relation graph encode a useful fragment of Sigma-style field conditions. This bridge places hand-crafted rule logic and learned relation masks in one representational framework, while remaining limited to single-email, field-level conditions.
    \item On a corpus of \dsTotalRecords{} emails, PhishSigma++ attains competitive clean F$_1$ scores and retains its detection accuracy under text‑padding attacks that cause substantial performance degradation in token‑based baselines.
\end{itemize}
\section{Background and Threat Model}\label{sec:background}
We focus on high-impact forms of social engineering, including credential phishing, business email compromise (BEC) style impersonation, fake invoices or delivery notices, and account-verification scams. These messages are designed to trigger sensitive user actions such as clicking a link, disclosing credentials, or transferring funds~\cite{das2019sok,cidon2019high}. In enterprise environments, such attacks are often low-volume and tailored to specific recipients rather than broadcast at campaign scale, reflecting the attacker’s goal of achieving a single successful outcome instead of maximizing delivery volume~\cite{verizon2023dbir,ho2019detecting}.
This distinction is important because indicators of bulk spam, including template reuse or repeated hits across many recipients, become weak signals when the attacker customizes the message for only one organization or employee. In contrast, studies of BEC and lateral phishing emphasize elements such as impersonated business roles, exploitation of workflow trust, and inconsistencies across fields that relate sender identity, recipient expectations, and linked resources~\cite{cidon2019high,ho2019detecting,gascon2018reading}. These are precisely the cues preserved by typed entity relations.
\begin{lstlisting}[style=emailstyle,float=t,caption={An anonymized credential-phishing email with displayed-link deception, urgency pressure, and brand–sender mismatch.},label={lst:background-email}]
From: "HarborView IT Service Desk" notice@account-review-mail.malicious
Reply-To: reset@account-review-mail.malicious
To: "Alice" alice@harborview-capital.target
Subject: Action Required: HarborView Mailbox Verification
<html><body> <p>Your <b>HarborView Capital</b> mailbox will be restricted unless you confirm your settings before today's cutoff.</p> <p><a href="https://account-review-mail.malicious/harborview/verify"> https://mail.harborview-capital.target/mailbox-review</a></p> <p>Regards,<br/>HarborView IT Service Desk</p> </body></html> \end{lstlisting}
In Listing~\ref{lst:background-email}, no single field is malicious on its own. The urgency phrasing is believable, the visible URL resembles an internal corporate link, and the brand name appears legitimate. Credential phishing succeeds by making local cues appear routine while the overall message contradicts the organization-specific context that the recipient expects.
What exposes the message is the inconsistency across typed entities. The brand string appears in both the display name and the body, yet the domains in the \texttt{From} and \texttt{Reply-To} fields do not belong to HarborView. The displayed URL resembles a corporate address but actually resolves to \texttt{account-review-mail.malicious}, which is controlled by the attacker. The deadline cue then encourages the recipient to follow that deceptive link. Traditional token-based or reputation-based filters such as SpamBayes and SpamAssassin~\cite{spambayes,spamassassin} fail to capture this pattern. The message has no attachment, targets only one user, and uses wording that may be novel. An attacker can further modify or pad the prose without altering the underlying social engineering goal, as illustrated by the Good Word attack against spam filters~\cite{lowd2005good}. Prior analyses of BEC and lateral phishing reach similar conclusions: sender identity, role expectations, and cross-field consistency are more reliable indicators than isolated textual features~\cite{cidon2019high,ho2019detecting}.
\paragraph{Threat model.}
The defender analyzes RFC822 emails with parseable headers at the mail gateway before user interface rendering, without access to external threat intelligence, DNS reputation services, or URL sandbox results. The attacker controls the subject line, body text, display name, embedded URLs, and certain mail user agent fields including \texttt{Reply-To} and \texttt{X-Mailer}, and can revise or pad body text at low cost. The attacker does not control DKIM, SPF, or DMARC authentication outcomes, cannot modify relay-inserted \texttt{Received} headers, and cannot eliminate payload-carrying URLs without abandoning the phishing objective. This boundary motivates separating attacker-written prose from more stable header, routing, and URL evidence. The robustness experiment in Section~\ref{sec:adversarial} implements a non-adaptive Good Word padding attack within this boundary. More advanced attacks such as adaptive header forgery, authenticated-domain abuse, or landing-page manipulation remain outside the scope of this work.
\section{Methodology}\label{sec:method}
\begin{figure}[t]
\centering
\includegraphics[width=\textwidth]{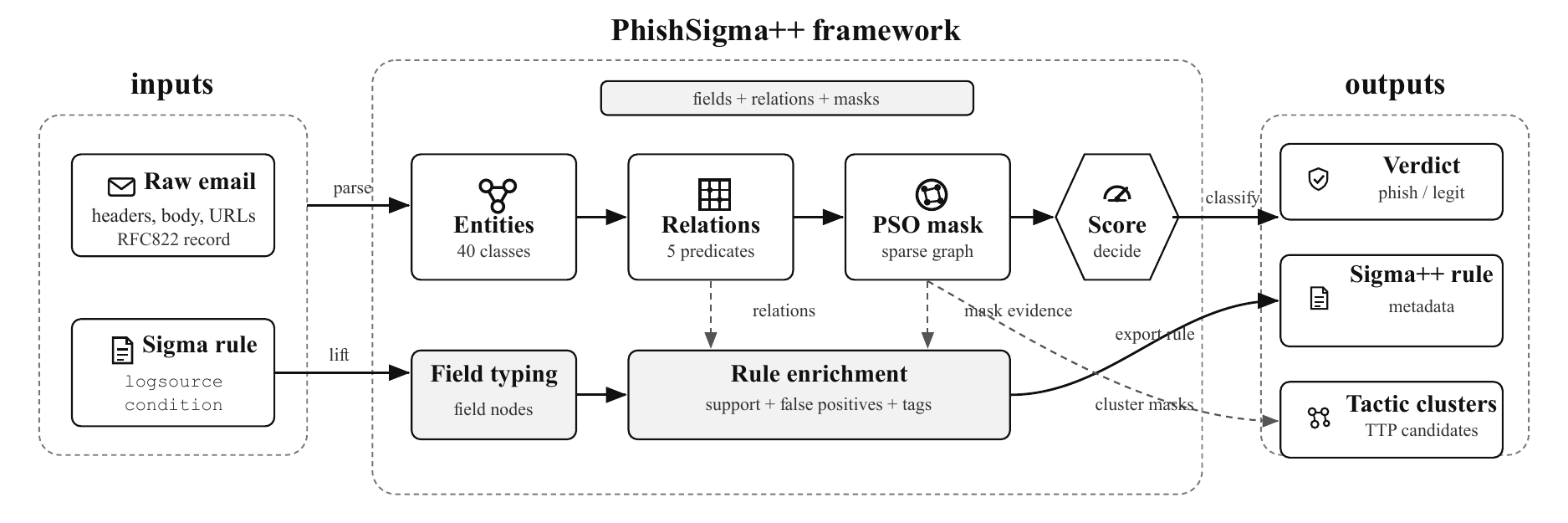}
\caption{PhishSigma++ system architecture.}
\label{fig:pipeline}
\end{figure}

Figure~\ref{fig:pipeline} shows the PhishSigma++ system architecture. The system consists of one detection pipeline and two parallel analysis outputs. The detection pipeline parses a raw RFC822 email into typed entities, scores their pairwise relations, selects a sparse PSO mask, and classifies the masked representation with a linear SVM to produce the final verdict. The same retained mask is then reused for analysis: thresholded relation scores are enriched into Sigma‑style rule metadata (support, false‑positive estimates, and tactic tags), and PSO‑pruned malicious vectors are clustered into candidate TTP families for review.
\subsection{Typed Entity-Relation Representation}\label{sec:entity-graph}
PhishSigma++ constructs the typed entity--relation graph from \entityLabels{} deterministic extractors (Appendix Table~\ref{tab:extractors}). Appendix Table~\ref{tab:extractors} groups them into Structural, Content, and Keyword extractors. The Structural extractors canonicalize header fields and decompose URLs into actual targets, displayed text, host-level components, subdomains, and query parameters; the Content group preserves subject and body spans; and the Keyword extractors assign textual spans to predefined phishing-related lexical categories. Importantly, extractor outputs are treated as typed candidate entities rather than fixed feature decisions. When a string admits multiple interpretations, PhishSigma++ preserves the corresponding typed occurrences, leaving PSO to select the entity occurrences and cross-type relations that contribute most to phishing detection.
\subsubsection{Abstract Relation Functions}
\begin{figure}[t]
\centering
\includegraphics[width=0.85\textwidth]{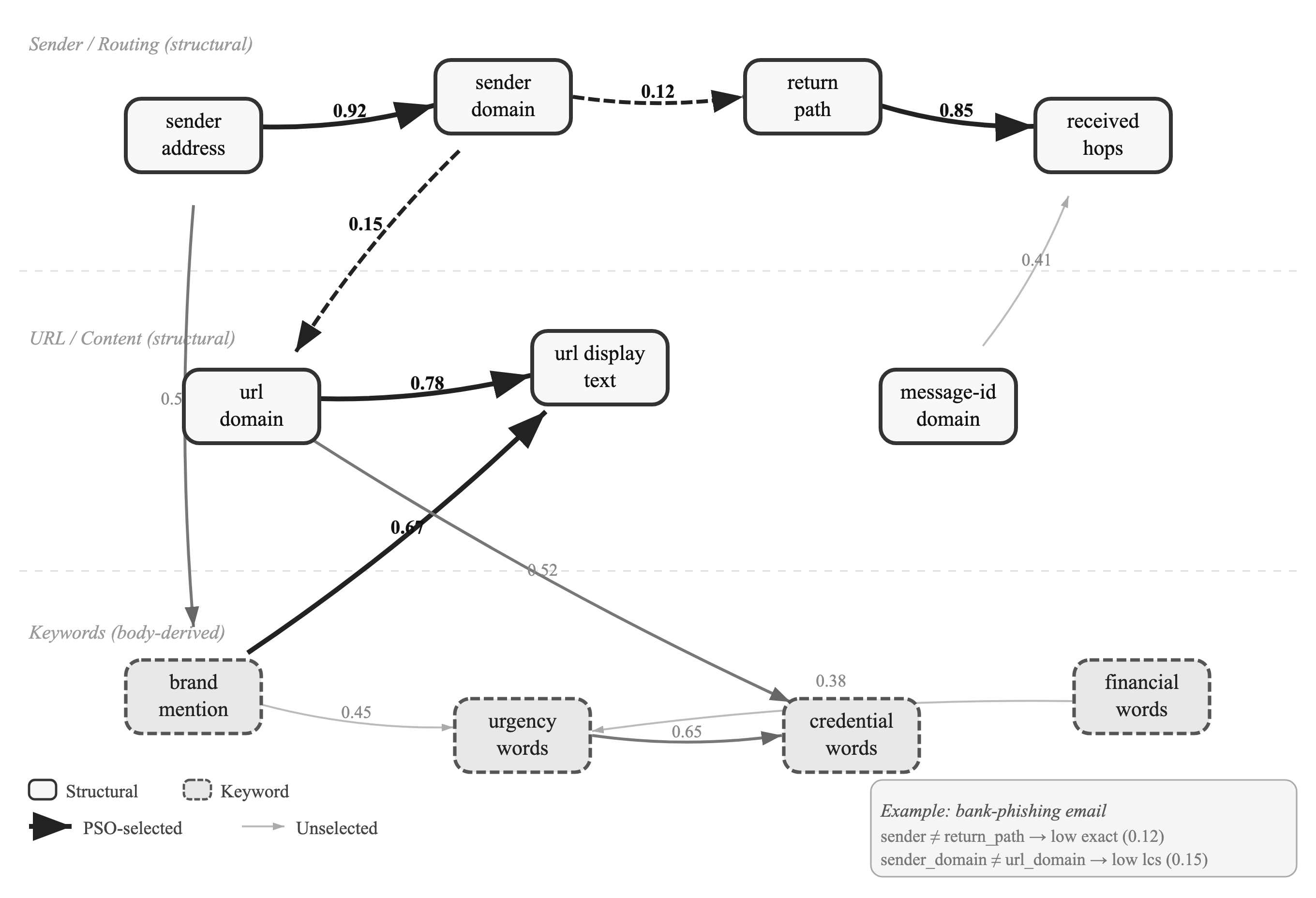}
\caption{Entity relation graph excerpt.}
\label{fig:graph-example}
\end{figure}

For each email~$e$, entities $v_1, \dots, v_k$ are extracted and pairwise relationships are computed using \numRelations{} functions returning scores in $[0,1]$:
\begingroup
\setlength{\jot}{2pt}
\begin{align}
f_{\mathrm{exact}}(s,t)
  &= \mathbb{1}\{s=t\},
  \label{eq:fexact} \\
f_{\mathrm{contain}}(s,t)
  &= \mathbb{1}\{s \subseteq t\}\,\frac{|s|}{|t|},
  \label{eq:fcontain} \\
f_{\mathrm{lcs}}(s,t)
  &= \max\!\left\{\frac{\operatorname{LCSubstr}(s,t)}{|s|},\, \frac{\operatorname{LCSubstr}(s,t)}{|t|}\right\},
  \label{eq:flcs} \\
f_{\mathrm{jaccard}}(s,t)
  &= \frac{|\operatorname{tok}(s) \cap \operatorname{tok}(t)|}{|\operatorname{tok}(s) \cup \operatorname{tok}(t)|},
  \label{eq:fjaccard} \\
f_{\mathrm{lenratio}}(s,t)
  &= \frac{\min(|s|,|t|)}{\max(|s|,|t|)}.
  \label{eq:flenratio}
\end{align}
\endgroup
These cover the qualitatively distinct similarity regimes that recur in header/URL/keyword comparisons: equality, inclusion, character-level overlap, token-level overlap, and length-shape equivalence. Adding finer-grained variants would be subsumed by the max-collapse below; an ablation that keeps all five functions as separate channels achieves $0.9639$ F$_1$ versus $0.9675$ for the collapsed mask (Section~\ref{sec:overall}), so the projection is preferred because it produces a single edge weight per type pair. The edge weight between types $\ell_i$ and $\ell_j$ is
\begin{equation}\label{eq:edge}
  w(\ell_i, \ell_j) = \max_r \max_{\substack{v_a : \ell(v_a)=\ell_i \\
    v_b : \ell(v_b)=\ell_j}} f_r(\text{str}(v_a), \text{str}(v_b)),
\end{equation}
yielding a directed weighted graph $G_e = (V_e, E_e, w)$ per email. Figure~\ref{fig:graph-example} shows an active subset: dark edges are PSO-selected, light dashed edges are unselected candidates, dashed node borders mark body-derived entities.
\subsubsection{Collapsed Matrix Representation}\label{sec:features}
Each graph is represented by $F(G) \in \mathbb{R}^{N \times N}$ with $N = \entityLabels{}$ (Appendix Table~\ref{tab:extractors}), flattened row-wise to $x(G) \in \mathbb{R}^{d}$, $d = N^2 = \featureDim$. Off-diagonal entries are $F_{i,j}(G) = w(\ell_i, \ell_j)$; the diagonal encodes presence, $F_{i,i}(G) = \min(c_i/5, 1)$ for match count $c_i$. The collapse from a $N \times N \times \numRelations{}$ tensor to $N \times N$ is a deterministic projection: relation-specific contrasts are discarded in exchange for a single edge weight per type pair, and any linear classifier on $x(G)$ is equivalent to the same classifier applied after this projection.
\subsection{Sparse Mask Selection}\label{sec:pso}
Exhaustive search over $2^d$ binary masks on the \featureDim{}-dimensional collapsed feature space is intractable~\cite{xue2015survey}. We use particle swarm optimization (PSO) with a real-valued mask $m \in [0,1]^d$; diagonal entries gate entity-type presence and off-diagonal entries gate cross-type edges. The mask remains continuous during search and is thresholded at $0.5$ only when rendering analyst summaries.
The fitness combines an inner cross-validated F$_1$ with a sparsity penalty:
\begin{equation}\label{eq:fitness}
  O(m) = \text{F}_1\bigl(\text{SVM}(\text{Prune}(\mathbf{X}, m),\;
  \mathbf{y})\bigr) - \lambda \cdot \frac{\|\text{active}(m)\|_0}{d},
\end{equation}
where $\text{Prune}(\mathbf{X}, m) = \mathbf{X} \odot m$ is row-broadcast element-wise gating, $\|\text{active}(m)\|_0$ counts coordinates above the $0.5$ threshold, and $\lambda = \psoSparsityLambda{}$. MaxAbsScaler is used because the features are highly sparse, so StandardScaler would distort zero-valued dimensions. The inner CV is kept lightweight because PSO performs roughly $\psoParticles{}\times\psoIters{}\times\totalRuns{}$ fitness evaluations; to prevent test-set leakage, it runs on a fixed PSO subset of \maxTrainPSO{} training emails sampled before the outer split and excluded from the outer test partition. Final results use the repeated evaluation protocol of Section~\ref{sec:eval}.
We use \psoParticles{} particles for \psoIters{} iterations with $\omega=0.7$, $c_1 = c_2 = 1.5$, and sparse initialization (off-diagonal $\sim U(0,0.35)$, diagonal $\sim U(0.3,0.8)$). PSO fits this problem because the objective is non-differentiable and multimodal in $m$, so gradient methods such as L1-SVM cannot be applied without first relaxing the sparsity term~\cite{xue2015survey}.
\begin{algorithm}[t]
\caption{PSO Optimization over the Collapsed Feature Space}\label{alg:pso}
\begin{algorithmic}[1]
\REQUIRE Collapsed features $\mathbf{X} \in \mathbb{R}^{n \times d}$, labels $\mathbf{y}$, where $d = N^2 = \featureDim$
\ENSURE Best-scoring collapsed mask $\mathbf{m}^{*} \in [0,1]^d$
\STATE Initialize particle masks $\mathbf{m}^{(p)}$ and velocities $\mathbf{v}^{(p)}$ with sparse priors
\FOR{$t = 1$ \TO \psoIters{}}
  \FOR{each particle $p$}
    \STATE $\mathbf{X}^{\text{prune}} \gets \text{Prune}(\mathbf{X}, \mathbf{m}^{(p)})$ \COMMENT{$\mathbf{X} \odot \mathbf{m}^{(p)}$, row-broadcast}
    \STATE Fit MaxAbsScaler on $\mathbf{X}^{\text{prune}}$; train LinearSVC
    \STATE $O^{(p)} \gets \text{F}_1^{\text{inner-CV}} - \lambda \cdot \text{sparsity}(\mathbf{m}^{(p)})$
    \STATE Update personal best $\mathbf{p}_{\text{best}}^{(p)}$
  \ENDFOR
  \STATE Update global best $\mathbf{m}^{*}$
  \FOR{each particle $p$}
    \STATE Draw $r_1, r_2 \sim U(0,1)$ \COMMENT{independent per dimension}
    \STATE $\mathbf{v}^{(p)} \!\gets\! \omega \mathbf{v}^{(p)} \!+\! c_1 r_1 (\mathbf{p}_{\text{best}}^{(p)} \!-\! \mathbf{m}^{(p)}) \!+\! c_2 r_2 (\mathbf{m}^{*} \!-\! \mathbf{m}^{(p)})$
    \STATE $\mathbf{m}^{(p)} \gets \text{clip}(\mathbf{m}^{(p)} + \mathbf{v}^{(p)},\, 0,\, 1)$
  \ENDFOR
\ENDFOR
\RETURN $\mathbf{m}^{*}$
\end{algorithmic}
\end{algorithm}
The output mask is used throughout the rest of the paper, reshaped into its $N\times N$ form for analyst-facing summaries.
\subsection{Detection and Analyst Outputs}\label{sec:classification}
PhishSigma++ applies the PSO-selected relation mask over header/URL entities and body-derived keyword categories, rescales sparse graph features with MaxAbsScaler, and trains a linear SVM. It therefore excludes raw bag-of-words text, but it does include deterministic body-derived semantic categories. Other learners are retained only as comparison, ablation, or robustness audits.
For analyst use, the same retained mask supports two secondary outputs. First, diagonal entries behave like IOC-style entity-presence checks and thresholded off-diagonal entries yield Sigma-compatible typed field conjunctions with support, false-positive estimates, and tactic tags. Section~\ref{sec:sigma-reduction} formalizes how this typed-rule view bridges to a Sigma-style field-condition fragment in the collapsed relation space considered in this paper. The current system therefore surfaces retained entity types and cross-type themes, not concrete field-value contradictions such as one specific sender domain conflicting with one specific URL domain. Second, we apply K-Means to PSO-pruned malicious feature vectors, choose $K$ by silhouette over $[2,10]$, and interpret each cluster centroid as a tactic template. Pairwise Pearson correlations within each cluster reveal which entity-type relationships co-vary within the same pattern. These analysis outputs explain the detector; they are not treated as standalone signatures.
\paragraph{Sigma-style relation-rule view.}\label{sec:sigma-reduction}
The bridge to Sigma-style logic is limited to single-email, field-level conditions over collapsed typed-relation scores. Let the pre-collapse tensor be $T(G) \in [0,1]^{N \times N \times R}$ with $T_{ijr}(G)=\phi_r(e_i,e_j)$, and let the deployed projection be $C(T)_{ij}=\max_{1 \le r \le R} T_{ijr}$ and $x(G)=\mathrm{vec}(C(T(G)))$. A handcrafted typed atom is then
\begin{equation}
  a_k(G)=\mathbb{1}[\sigma_k(x_k(G)-\tau_k) \ge 0], \qquad \sigma_k \in \{+1,-1\}.
\end{equation}
Diagonal atoms encode typed-entity presence and off-diagonal atoms encode thresholded typed relations. For a clause support $S_c \subseteq \{1,\dots,d\}$,
\begin{equation}
  C_{S_c}(G)=\bigwedge_{k \in S_c} a_k(G), \qquad
  f_{S_c}(G)=\sum_{k \in S_c} a_k(G)-|S_c|+\tfrac{1}{2}.
\end{equation}
Because the atoms are binary, $f_{S_c}(G)>0$ holds exactly when all atoms in the clause fire. Finite disjunctions of such clauses therefore recover a useful collapsed Sigma-style fragment, and the deployed PSO+SVM model strictly generalizes this rule mode by learning the support and weights from data.
Two formal boundaries matter. First, max-collapse is not invertible for $R \ge 2$: two tensors that swap the maximizing relation channel can induce the same collapsed edge while disagreeing on a relation-specific atom. Second, if the full tensor is retained, the rule language is strictly richer. Every collapsed lower-threshold atom can be rewritten as $\bigvee_{r=1}^{R} \mathbb{1}[T_{ijr} \ge \tau]$, and every collapsed upper-threshold atom as $\bigwedge_{r=1}^{R} \mathbb{1}[T_{ijr} \le \tau]$; however, channel-specific tests cannot in general be recovered from $C(T)$ alone. Hence removing max-collapse strictly increases relation-rule coverage. Within the deployed classifier $s(G)=w^\top x(G)+b$, any two emails with the same $x(G)$ receive the same score, so the current max-collapse changes explanation granularity rather than the decision surface of the deployed model.
\paragraph{Concrete mapping instance.}
Listing~\ref{lst:background-email} contains a brand--sender mismatch: the message presents the HarborView brand while the sender domain is \texttt{account-review-mail.malicious}. In the current implementation, the \entityLabels{} extractor identifiers are sorted lexicographically before row-wise flattening, so \texttt{brand\_mentions} is index $i=5$ and \texttt{sender\_domain} is index $j=28$. The corresponding collapsed coordinate is therefore $k=(i-1)N+j=(5-1)\cdot 40 + 28 = 188$; this index is deterministic once the extractor vocabulary and its lexicographic ordering are fixed. For Listing~\ref{lst:background-email}, the worked example gives $x_{188}(G)=w(\text{brand\_mentions},\text{sender\_domain})\approx 0.05$. A handcrafted mismatch rule is therefore just the threshold test $x_{188}(G) \le 0.1$, i.e., one collapsed coordinate plus one cutoff. This is the concrete sense in which a field-consistency rule becomes a typed relation test in the deployed feature space.
\paragraph{Learned versus handcrafted masks.}
The reduction above is representational: it shows that a handcrafted Sigma-style mask can be embedded exactly in the typed-relation space. It does \emph{not} imply that the PSO-learned support must numerically coincide with a human-authored Sigma rule base. In practical terms, constructing a large, high-quality manual Sigma corpus for RFC822 malicious-email detection would require substantial expert effort, and this paper does not include such a corpus as a direct baseline; the Literal IOC/rule row in Table~\ref{tab:ablation} is therefore only an external literal-rule baseline, not a full handcrafted Sigma equivalent. The practical value of the generalization claim is that PSO searches a \featureDim{}-dimensional typed-relation space and can recover structured cross-field supports far beyond the few dozen rules that human analysts would ordinarily enumerate manually. The learned mask should therefore be read as data-driven rule discovery over the same representational substrate, rather than as evidence that a small handcrafted Sigma mask and the learned support are identical. Because the current implementation uses the max-collapse of Eq.~(\ref{eq:edge}), it does not preserve which relation function achieved the score. The formal propositions above show that this limits explanation recovery and channel-specific Sigma coverage, while leaving the deployed collapsed classifier unchanged on any pair of emails with the same $x(G)$. The method still does not cover multi-event correlation, list lookups, or aggregation operators from the full Sigma language.
\section{Evaluation}\label{sec:eval}
\subsection{Dataset and Setup}\label{sec:dataset}
\begin{figure}[t]
\centering
\includegraphics[width=0.78\textwidth]{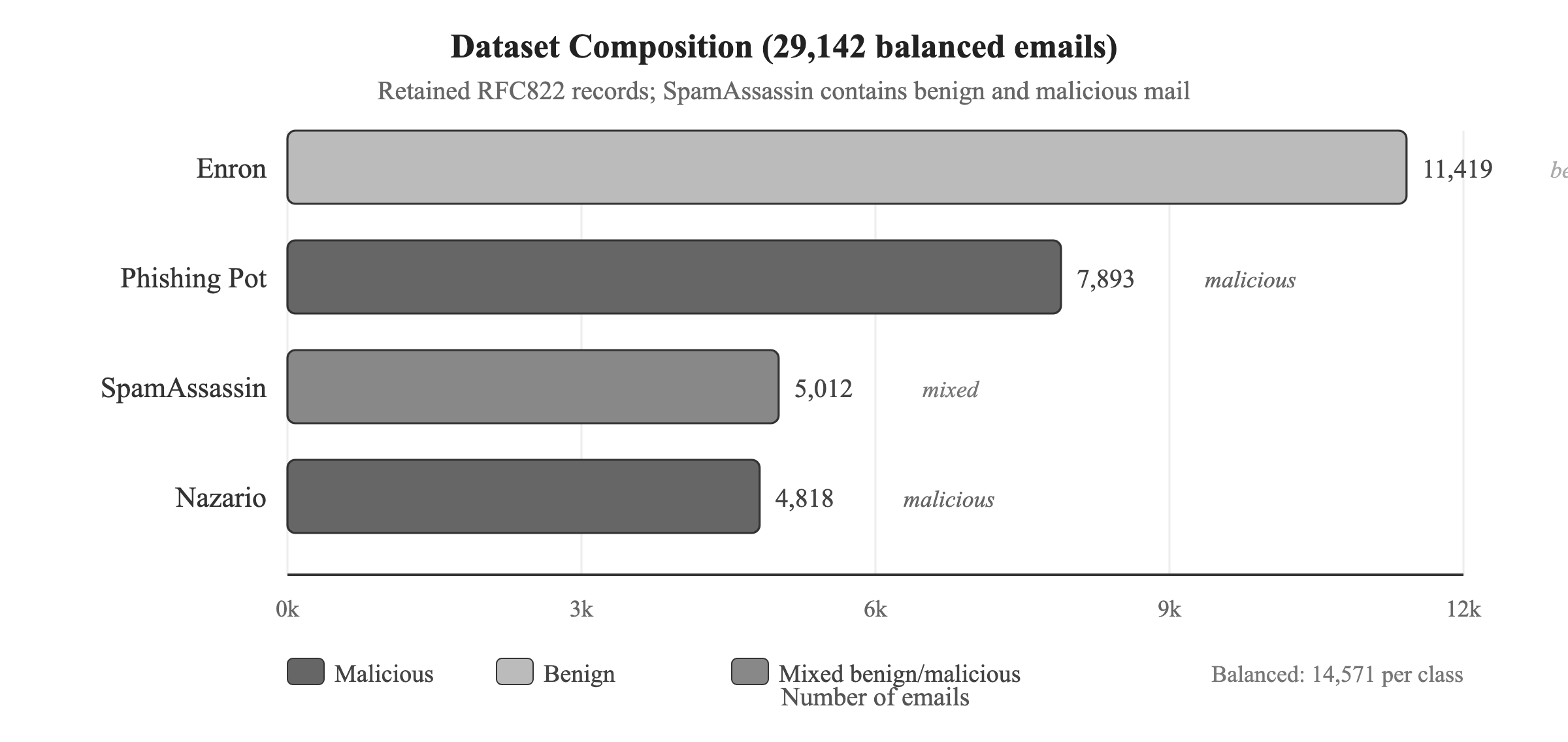}
\caption{Dataset composition under the binary evaluation setting.}
\label{fig:dataset}
\end{figure}

We evaluate on a class-balanced set of \dsTotalRecords{} RFC822 emails (\dsPhishCount{} positive, \dsLegitCount{} benign) drawn from three public corpora and one public repository snapshot (Fig.~\ref{fig:dataset}): Enron~\cite{klimt2004enron} (11{,}419 benign), Phishing Pot~\cite{phishingpot2024} (7{,}893 phishing), Nazario~\cite{nazario2006phishingcorpus} (4{,}818 phishing), and SpamAssassin~\cite{spamassassin} (3{,}152 ham, 1{,}860 spam). Phishing Pot does not publish a formal tagged release; we therefore cite the repository snapshot accessed on 2025-01-15 and ingest the upstream \texttt{email/*.eml} files as distributed. Our loader applies only RFC822 validity checks (minimum header presence and non-trivial message length) before feature extraction; it does not manually rewrite or relabel the sample content. SpamAssassin spam is mapped to the positive class because its urgency, impersonation, delivery-deception, and credential/payment-lure patterns would otherwise contaminate the benign side; Table~\ref{tab:split-clean} confirms that the spam slice is harder than the phishing-only slices, so the positive class is heterogeneous and reported metrics describe binary malicious-email detection rather than a phishing-only benchmark. Header coverage is \headerCoverageFrom{} (\texttt{From}), \headerCoverageSubject{} (\texttt{Subject}), \headerCoverageMessageId{} (\texttt{Message-ID}); entity extraction covers \entityCoverageAll{} of records (avg \avgEntityTypes{} types/email). All supervised results use repeated stratified cross-validation across \numSeeds{} seeds (\totalRuns{} runs)~\cite{arp2022and}.
\subsection{Detection Results}\label{sec:overall}
\setcounter{table}{1}
\begin{table}[t]
\centering
\caption{Clean detection results on the balanced \dsTotalRecords{}-email corpus; bold marks the best F$_1$.}
\label{tab:overall}
\footnotesize
\setlength{\tabcolsep}{2.5pt}
\begin{tabular}{@{}p{0.46\linewidth}cccc@{}}
\toprule
\textbf{Method} & \textbf{F$_1$} & \textbf{P} & \textbf{R} & \textbf{Acc.} \\
\midrule
RuleIOC (2020)~\cite{iocfinder2020} & \ruleIocFOne{} & \ruleIocPrec{} & \ruleIocRec{} & 0.532 \\
SpamBayes (2011)~\cite{spambayes} & \spamBayesFOne{} & \textbf{\spamBayesPrec{}} & \spamBayesRec{} & 0.960 \\
SpamAssassin~\cite{spamassassin} & 0.1583 & 0.8739 & 0.0870 & 0.537 \\
Dima806 DistilBERT (2023)\footnotemark[1] & 0.9367 & 0.9356 & 0.9379 & 0.937 \\
ElSlay BERT (2024)\footnotemark[2] & 0.8584 & 0.9565 & 0.7786 & 0.872 \\
E-PhishGen NB (2025)~\cite{pajola2025phishgen} & 0.8692 & 0.9793 & 0.7840 & 0.884 \\
\textbf{PhishSigma++} & \textbf{\ppFOne{}} & \ppPrec{} & \textbf{\ppRec{}} & \textbf{0.968} \\
\bottomrule
\end{tabular}
\end{table}
\footnotetext[1]{\url{https://huggingface.co/dima806/phishing-email-detection}}
\footnotetext[2]{\url{https://huggingface.co/ElSlay/BERT-Phishing-Email-Model}}
Table~\ref{tab:overall} compares PhishSigma++ against rule-, token-, and content-based baselines on the same \dsTotalRecords{}-email corpus. Rows with local re-training report aggregate results over their own resampling protocols: PhishSigma++ uses the main repeated evaluation protocol, and the E-PhishGen NB row re-fits the released TF-IDF+NB adapter design on the same balanced local RFC822 corpus. The Dima806 and ElSlay rows are released public checkpoints applied directly to our corpus as transfer baselines. This choice is deliberate: the main comparison target in this paper is not best-case clean accuracy after local re-tuning, but the amount of performance lost once the same released model is exposed to body-text perturbation. Using the checkpoints directly keeps that robustness-loss comparison fair, because local fine-tuning would conflate robustness with paper-specific optimization, data-balancing, and early-stopping choices. SpamAssassin and SpamBayes are single full-corpus runs. PhishSigma++ reaches \ppFOne{} F$_1$ from typed entity relations and keyword categories alone, without unrestricted bag-of-words text, and trades only $\Delta\text{F}_1\!\approx\!0.009$ against SpamBayes on this balanced corpus. The clean numbers compress because token-based classifiers operating on body text are already near the corpus ceiling; the discriminating comparison is the body-padding stress of Section~\ref{sec:adversarial}, where the same configurations diverge by more than $30\times$. Within the clean setting, PhishSigma++ also yields the highest accuracy in Table~\ref{tab:overall} while preserving typed, analyst-reviewable evidence rather than unrestricted lexical features.
\begin{table}[t]
\centering
\caption{Source-wise slice accuracy at $\rho=0.0$.}
\label{tab:split-clean}
\footnotesize
\setlength{\tabcolsep}{3pt}
\begin{tabular}{@{}p{0.40\linewidth}p{0.24\linewidth}c@{}}
\toprule
\textbf{Dataset} & \textbf{Category} & \textbf{Slice Acc.} \\
\midrule
Nazario & phishing-only (+) & 0.9907 \\
Phishing Pot & phishing-only (+) & 0.9890 \\
SpamAssassin spam & spam-only (+) & 0.7337 \\
Enron & benign-only ($-$) & 0.9969 \\
SpamAssassin ham & benign-only ($-$) & 0.8905 \\
\bottomrule
\end{tabular}
\end{table}
Table~\ref{tab:split-clean} reports source-wise slice accuracy at $\rho=0.0$. SpamAssassin is harder than the other collections on \emph{both} sides of the boundary: its spam slice and its ham slice achieve the lowest source-specific accuracies, consistent with that collection being a topically distinct, older-style commercial- and bulk-mail corpus. The residual clean errors are therefore concentrated in SpamAssassin on both sides of the binary boundary.
\paragraph{Ablation.}
Table~\ref{tab:ablation} shows that literal IOC matching alone is weak at 0.2031 F$_1$, while typed entity counts already lift performance to 0.9372. Adding the PSO-pruned relation graph gives the best result, 0.9675 F$_1$, a further +0.0303 over counts alone. Keeping all interaction features is slightly worse at 0.9639, indicating that selective relation evidence is more useful than the full dense interaction set.
\begin{table}[t]
\centering
\caption{Component ablation.}
\label{tab:ablation}
\footnotesize
\setlength{\tabcolsep}{3pt}
\begin{tabular*}{\linewidth}{@{\extracolsep{\fill}}llcc@{}}
\toprule
\textbf{Configuration} & \textbf{Role in the design} & \textbf{F$_1$} & $\Delta$ \\
\midrule
Literal IOC/rule matching & External indicator coverage & 0.2031 & -- \\
Entity counts only & Typed entities without relations & 0.9372 & +0.7341 \\
PSO-pruned relation graph & PhishSigma++ detector & \textbf{0.9675} & +0.0303 \\
All interaction features & Dense relation control & 0.9639 & -0.0036 \\
\bottomrule
\end{tabular*}
\end{table}
\subsection{Adversarial Robustness}\label{sec:adversarial}
We evaluate robustness to a Good Word body-injection attack on the same full corpus as the main results. At injection rate $\rho$, phishing test emails are padded with text drawn from benign emails so that a fraction $\rho$ of the final body is benign; headers and URLs are preserved (MTA-controlled or required for the attack to succeed). After injection, structural features are re-extracted from the perturbed text, so any new entities introduced by the padding are represented in the evaluated features. Locally re-fit models are evaluated on the perturbed corpus and reported by F$_1$; released checkpoints and direct legacy filters are evaluated once on the same perturbed corpus with no additional training.
\begin{table}[t]
\centering
\caption{Good Word body-injection robustness measured by F$_1$.}
\label{tab:adversarial-main}
\footnotesize
\setlength{\tabcolsep}{2.5pt}
\begin{tabular*}{\linewidth}{@{\extracolsep{\fill}}lccccc@{}}
\toprule
\textbf{Method} & $\rho$=0.0 & $\rho$=0.2 & $\rho$=0.4 & $\rho$=0.6 & $\rho$=0.8 \\
\midrule
Dima806 DistilBERT (2023)\footnotemark[1] & 0.9367 & 0.8472 & 0.7395 & 0.7345 & 0.7284 \\
ElSlay BERT (2024)\footnotemark[2] & 0.8584 & 0.7930 & 0.7083 & 0.7003 & 0.6851 \\
E-PhishGen NB adapter (2025)~\cite{pajola2025phishgen} & 0.8692 & 0.8366 & 0.7925 & 0.6914 & 0.4956 \\
SpamBayes (2011)~\cite{spambayes} & 0.9672 & 0.8080 & 0.3299 & 0.0633 & 0.0243 \\
\textbf{PhishSigma++} & \textbf{0.9675} & \textbf{0.9642} & \textbf{0.9642} & \textbf{0.9612} & \textbf{0.9579} \\
\bottomrule
\end{tabular*}
\end{table}
\footnotetext[1]{\url{https://huggingface.co/dima806/phishing-email-detection}}
\footnotetext[2]{\url{https://huggingface.co/ElSlay/BERT-Phishing-Email-Model}}
Table~\ref{tab:adversarial-main} mixes three evaluation modes. PhishSigma++ and the E-PhishGen NB adapter are locally re-fit on the perturbed corpus, the Dima806 and ElSlay rows are released public checkpoints re-evaluated on perturbed local emails with no additional training, and SpamBayes is a deterministic direct stress run on the full perturbed corpus. This is the fairest setting for the robustness comparison in this paper: it exposes the same released model or released adapter design to the clean and perturbed corpora and measures the resulting degradation, rather than mixing attack sensitivity with paper-specific re-tuning. PhishSigma++ stays nearly flat from 0.9675 at $\rho=0.0$ to 0.9579 at $\rho=0.8$, whereas the E-PhishGen adapter drops from 0.8692 to 0.4956 over the same range. The text-heavy systems collapse under padding because their discriminative tokens are diluted; SpamBayes exhibits the textbook Good Word failure mode~\cite{lowd2005good}.
The robustness has a structural source. Good Word padding is applied \emph{before} extraction, so every feature in $F(G)$ is recomputed from the perturbed message; what stays stable is the feature distribution rather than a cached vector. Three properties matter. First, the Structural header and URL types, such as sender fields, reply-to, return-path, received hops, message-id domains, x-mailer fields, content-type fields, and URL components, are not modified by body insertions, so their diagonal entries and incident edges stay unchanged across $\rho$. Second, the Keyword extractors are lexicon-bounded and the diagonal saturates at $\min(c_i/5,1)$, so benign padding does not erase the original urgency, credential, or financial cues unless the padding itself introduces stronger malicious cues, which the protocol forbids. Third, the PSO mask concentrates weight on cross-type edges between Structural types and these stable Keyword categories (Table~\ref{tab:pso-nodes}), so the SVM input is dominated by features whose distribution is bounded by the attacker's permitted edits. Empirically, benign-side predictions remain unchanged across $\rho \in \{0,0.2,0.4,0.6,0.8\}$ because benign emails are not padded, and the positive-side recall drops by under one percentage point on the phishing slices while remaining in the same range on the SpamAssassin-spam slice. These structural properties hold only for non-adaptive padding within the threat model of \S\ref{sec:background}; adaptive header forgery, URL obfuscation, and entity splicing are explicitly out of scope and are listed in \S\ref{sec:limitations}.
\subsection{Evidence and Tactic Analysis}\label{sec:tactics-eval}
The retained mask supports an analyst-facing summary that uses the same parameters as the classifier, with no separate explainer model.
\begin{table*}[t]
\centering
\caption{PSO-selected tactic evidence (overlapping coverage).}
\label{tab:pso-nodes}
\footnotesize
\setlength{\tabcolsep}{3pt}
\begin{tabular}{@{}p{0.20\textwidth}p{0.27\textwidth}p{0.11\textwidth}p{0.34\textwidth}@{}}
\toprule
\textbf{Theme} & \textbf{Evidence} & \textbf{Coverage} & \textbf{Reading} \\
\midrule
Sender-URL divergence & sender identity with url structure & 66\% & Spoofed identity points away from the payload domain. \\
Urgent link funnel & urgency language with url structure & 37\% & Deadline language steers the user toward a link. \\
Infrastructure-payload coupling & routing metadata with url structure & 27\% & Routing traces recur with payload URL structure. \\
Authority-payment lure & authority mimicry with financial language & 20\% & Authority cues frame credential or payment requests. \\
Brand urgency lure & brand impersonation with urgency language & 17\% & Brand names pair with suspension or deadline pressure. \\
Brand-sender mismatch & brand impersonation with sender identity & 17\% & Content brand conflicts with sender identity fields. \\
\bottomrule
\end{tabular}
\end{table*}
Table~\ref{tab:pso-nodes} groups the most frequent retained relations into type-level themes (rows overlap, so coverage does not sum to 100\%); cross-seed Jaccard \psoJaccard{} indicates stability under random restarts.
\paragraph{Worked example.}
Running Listing~\ref{lst:background-email} through the deployed pipeline yields a brand-bearing display signal, \texttt{brand\_mentions}, with value ``HarborView Capital'', a sender domain that does not match the brand, a distinct reply-to domain, a URL domain with no shared tokens, and the keyword categories urgency, link-action, and credential. In the collapsed matrix, representative edges include $w(\text{sender\_domain},\text{url\_domain})\approx 0.18$ from length-ratio and $w(\text{brand\_mentions},\text{sender\_domain})\approx 0.05$ from token Jaccard; the urgency coordinate is also active with the URL domain. Among the \featureDim{} coordinates, the PSO mask retains only a small subset. The four largest positive SVM weights connect sender identity with URL structure, brand impersonation with sender identity, urgency with URL structure, and credential cues with URL structure. The analyst-facing summary therefore reads as four typed lines (sender--URL divergence, brand--sender mismatch, urgency--link pressure, credential--link pressure) rather than a token list, and the same mask subsequently groups the email into the ``brand urgency lure'' family of Appendix Table~\ref{tab:novel-patterns}. Because the summary is generated by the deployed model itself, a changed mask immediately changes the analyst-facing lines.
Across \totalRuns{} runs, K-means on PSO-pruned malicious vectors (per-run $K$ chosen by silhouette over $[2,10]$, centroid signatures unioned) yields \numTacticPatterns{} distinct cluster-centroid labels. Of \numUniqueTactics{} unique patterns, \numMitreCovered{} map directly to MITRE ATT\&CK~\cite{mitre2023attack} techniques and \numNovelTactics{} ($\pctNovelTactics{}\%$) are cross-field correlations not encoded at the technique level of ATT\&CK; this share reflects taxonomy granularity rather than a claim of newly discovered attacker techniques. Sender--URL divergence appears in two-thirds of families and brand--urgency pairing in about one-sixth. Some clusters are narrow and immediately actionable for analyst review, while broader clusters remain exploratory. Appendix Table~\ref{tab:novel-patterns} lists representative patterns without a direct ATT\&CK counterpart.
\section{Discussion}\label{sec:discussion}
\paragraph{Clean accuracy and robustness are separate axes.}
The strongest clean-data endpoints come from content-aware models, but those systems provide less direct typed-field provenance and degrade once body text is diluted. PhishSigma++ trades only $\Delta\text{F}_1\!\approx\!0.009$ versus SpamBayes for an order-of-magnitude smaller drop under body padding and analyst-facing evidence at the type level.
\paragraph{A sparse malicious-pattern detector, not a campaign signature.}
The heterogeneous positive class makes the learned mask a malicious-pattern detector rather than a phishing-campaign signature: shared cross-field structure transfers across phishing and spam slices, and the lower spam-only row in Table~\ref{tab:split-clean} reflects the same heterogeneity. Tactic clusters expose this mask at the type level but remain analyst-review candidates rather than validated threat-intelligence rules.
\paragraph{Cost separation between training and inference.}
PSO and tactic clustering are a single offline training stage (\totalTimeMin{} min on \dsTotalRecords{} emails) that runs at the cadence of model refreshes. Gateway inference uses the frozen extractors, frozen mask, and one linear-SVM decision function and processes \engineThroughput{} emails/s; the per-email cost is deterministic extraction plus one \featureDim{}-dimensional sparse dot product, and the offline search complexity affects refresh cadence rather than online throughput.
\section{Limitations and Future Work}\label{sec:limitations}
Regex extraction may not cover all encodings, the five relation functions capture string similarity rather than deeper semantics, max-collapse discards relation-specific contrasts; retaining the full relation tensor would let the mask select relation channels and inequality directions directly, bringing the implementation closer to richer Sigma-style field logic; and PSO is a metaheuristic (cross-seed Jaccard \psoJaccard{} suggests stability but not optimality). Stratified CV may overestimate generalization if source identity correlates with labels; the binary setup maps SpamAssassin spam to the positive class, so results are not phishing-only campaign generalization. The body-padding stress is non-adaptive: attacks that target the typed-relation surface itself --- adaptive URL obfuscation lowering the sender--URL edge weight, brand-token splicing into the display name, or benign-header splicing to dilute relation evidence --- require attacker-aware protocols and remain future work, alongside authenticated-header forgery, sender-reputation manipulation, landing-page rewriting, concept drift~\cite{pendlebury2019tesseract}, and larger pretrained text encoders.
\section{Related Work}\label{sec:related-work}
Email-phishing detection has largely been framed as a content problem. URL-lexical models~\cite{sahingoz2019machine}, deep-learning text classifiers~\cite{bountakas2023helphed,sanh2019distilbert}, and the broader survey of Das et al.~\cite{das2019sok} show that strong clean performance is achievable when the model can freely consume body text, URLs, and surface tokens. These systems are effective at recognizing recurring lexical patterns, but their evidence is often difficult to map back to the operational question an analyst asks about a specific message: which fields disagree, which role is being impersonated, and which link or sender relationship actually made the email suspicious.
A second line of work focuses on organizational phishing, where those cross-field inconsistencies matter more than broad text similarity. Cidon et al.~\cite{cidon2019high} characterize BEC through sender behavior and historical communication context; Ho et al.~\cite{ho2019detecting} detect lateral phishing from header-level and behavioral anomalies at enterprise scale; Gascon et al.~\cite{gascon2018reading} show that content-agnostic email features can separate spear-phishing from benign traffic. Together these works motivate a detector that treats phishing as abuse of business context rather than as generic spam text. Our setting is narrower than theirs in that we stay within single-message RFC822 evidence, but richer than pure header anomaly scoring because we explicitly model relations among header entities, brand cues, and embedded URLs.
Recent email-security studies further highlight why this distinction matters. Reporting-evasion techniques, user-facing URL-inspection tasks, and blocklist manipulation all exploit the gap between what the message looks like locally and what it is doing operationally~\cite{chand2025doubly,lain2025url,li2025hades}. Those works motivate email-specific evaluation and analyst-facing evidence, but they do not model pairwise relations between typed entities inside one message.
The clean-accuracy/robustness gap has been studied since Dalvi et al.~\cite{dalvi2004adversarial} and the Good Word attack~\cite{lowd2005good}; Li et al.~\cite{li2018textbugger} generalize the pattern to neural text systems and Biggio et al.~\cite{biggio2013evasion} study evasion against learned spam filters. These attacks target token-level or differentiable feature surfaces. Our typed-relation representation reduces that exposure for the specific non-adaptive setting we test, because header and URL extractors remain stable under simple body-text padding even when the prose itself changes.
Intelligence-driven methods anchor detection on curated knowledge. IOC extractors~\cite{iocfinder2020} and CTI aggregation~\cite{bouwman2020different} treat indicators largely in isolation. Sigma rules~\cite{sigma2023spec,gao2021enabling} encode analyst-written SIEM detections as structured YAML rules. Graph-based provenance systems~\cite{han2020unicorn,milajerdi2019holmes,hassan2019nodoze} capture rich relational structure, but they do so at the system-call or network level. PhishSigma++ instead applies PSO-guided graph masks \emph{within} single emails~\cite{xue2015survey}, so the same sparse typed-relation structure drives both the verdict and the analyst-reviewable evidence summary.
\section{Conclusion}\label{sec:conclusion}
PhishSigma++ couples a typed entity-relation graph with a PSO-selected sparse mask to produce both a malicious-email verdict and type-level relation evidence from RFC822 messages. On \dsTotalRecords{} emails it reaches \ppFOne{} F$_1$ on clean data and retains \ppFOneAdvHigh{} F$_1$ under Good Word body padding at $\rho=0.8$, where token-based baselines drop by an order of magnitude. The same retained mask also groups malicious examples into \numTacticPatterns{} candidate tactic families, showing that detection accuracy, evidence provenance, and robustness should be evaluated together.
\bibliographystyle{unsrt}
\bibliography{references}
\clearpage
\appendix
\renewcommand{\thetable}{A\arabic{table}}
\setcounter{table}{0}
\section{Supplementary Tables}\label{app:tables}
\begin{table}[H]
\centering
\caption{Entity extractor groups.}
\label{tab:extractors}
\footnotesize
\setlength{\tabcolsep}{3pt}
\begin{tabular*}{\textwidth}{@{\extracolsep{\fill}}p{0.23\textwidth}p{0.12\textwidth}p{0.52\textwidth}c@{}}
\toprule
\textbf{Extractor} & \textbf{Group} & \textbf{Entity Types} & \textbf{Count} \\
\midrule
Sender identity & Structural & sender\_address, display\_name, domain, return\_path, reply\_to & 5 \\
Routing & Structural & received\_hops, x\_mailer, message\_id\_domain, content\_type & 4 \\
Recipient & Structural & recipient\_address, domain, username & 3 \\
URL structure & Structural & url\_actual, display\_text, domain, subdomain, params & 5 \\
Attachment & Structural & attachment\_info & 1 \\
Content & Content & subject\_text, body\_text & 2 \\
Direct indicators & Keyword & urgency, action, threat, financial, credential & 5 \\
Brand / social eng. & Keyword & brand, greeting, closing, link-action, time-pressure, legal-threat & 6 \\
Authority / scare & Keyword & government, notification, tech-jargon, scam-prize, security-alert, account-status & 6 \\
Financial deception & Keyword & money-transfer, personal-info, shipping & 3 \\
\bottomrule
\end{tabular*}
\end{table}
\clearpage
\begin{table}[H]
\centering
\caption{Representative cross-entity patterns without direct MITRE ATT\&CK counterparts.}
\label{tab:novel-patterns}
\footnotesize
\setlength{\tabcolsep}{2pt}
\begin{tabularx}{\textwidth}{@{}>{\raggedright\arraybackslash}p{0.23\textwidth}>{\raggedright\arraybackslash}p{0.20\textwidth}>{\raggedright\arraybackslash}X>{\raggedleft\arraybackslash}p{0.07\textwidth}@{}}
\toprule
\textbf{Cross-Type Pattern} & \textbf{Entity Groups} & \textbf{Why Not in MITRE} & \textbf{\#Tactics} \\
\midrule
  Sender--URL domain divergence & sender identity $\leftrightarrow$ URL structure & MITRE separates spearphishing links from masquerading rather than encoding one cross-field relation. & 27 \\
  Urgency--link pressure pattern & urgency language $\leftrightarrow$ URL structure & No MITRE technique models urgency language correlated with a linked action target. & 15 \\
  Infrastructure--payload fingerprint & routing metadata $\leftrightarrow$ URL structure & MITRE does not model sending infrastructure and payload-URL correlation as a distinct pattern. & 11 \\
  Authority--financial exploitation & authority mimicry $\leftrightarrow$ financial language & No MITRE technique models authority impersonation co-occurring with financial-request language. & 8 \\
  Brand impersonation + urgency pressure & brand cues $\leftrightarrow$ urgency language & No single MITRE technique captures the brand-urgency co-occurrence. & 7 \\
  Brand--sender identity mismatch & brand cues $\leftrightarrow$ sender identity & T1036.005 covers name matching, not brand-sender cross-field correlation. & 7 \\
  Fear--link exploitation & fear cues $\leftrightarrow$ URL structure & Social engineering is generic in MITRE; fear-to-link coupling is outside its granularity. & 4 \\
\bottomrule
\end{tabularx}
\end{table}
\end{document}